\documentclass[pra,preprintnumberssuperscriptaddress,nofootinbib]{revtex4-1}  
\usepackage{graphicx}
\usepackage{amsmath}
\usepackage{bm}

\def\ket#1{\left\vert #1 \right\rangle}
\def\bra#1{\left\langle #1 \right\vert}
\def\braket#1#2{\left\langle #1 \vert #2\right\rangle}

\newcommand\C{\hbox{$\mit I$\kern-.7em$\mit C$}}
\newcommand\R{\hbox{$\mit I$\kern-.6em$\mit R$}}

\begin{document}

\title{Quantum Algorithms for Quantum Chemistry based on the sparsity of the CI-matrix} 
\author{Borzu Toloui}%
\email{borzumehr@gmail.com}
\affiliation {Department of Physics and Astronomy, Haverford College, PA, 19041, USA.}
\author{Peter Love}
\affiliation {Department of Physics and Astronomy, Haverford College, PA, 19041, USA.}

\begin{abstract}
Quantum chemistry provides a target for quantum simulation of considerable scientific interest and industrial importance. The majority of algorithms to date have been based on a second-quantized representation of the electronic structure Hamiltonian - necessitating qubit requirements that scale linearly with the number of orbitals. The scaling of the number of gates for such methods, while polynomial, presents some serious experimental challenges. However, because the number of electrons is a good quantum number for the electronic structure problem it is unnecessary to store the full Fock space of the orbitals. Representation of the wave function in a basis of Slater determinants for fixed electron number suffices. However, to date techniques for the quantum simulation of the Hamiltonian represented in this basis - the CI-matrix - have been lacking. We show how to apply techniques developed for the simulation of sparse Hamiltonians to the CI-matrix. We prove a number of results exploiting the structure of the CI-matrix, arising from the Slater rules which define it, to improve the application of sparse Hamiltonian simulation techniques in this case. We show that it is possible to use the minimal number of qubits to represent the wavefunction, and that these methods can offer improved scaling in the number of gates required in the limit of fixed electron number and increasing basis set size relevant for high-accuracy calculations. We hope these results open the door to further investigation of sparse Hamiltonian simulation techniques in the context of the quantum simulation of quantum chemistry.
\end{abstract}

\maketitle 

\section{Introduction}
\label{sec:Intro}
Quantum computation offers the promise of large algorithmic improvements over classical calculations for specific problems. Algorithmic speedups are known for quantum simulation, quantum search, and algorithms (including Shor's factoring algorithm) based on hidden subgroup problems and related techniques~\cite{Grover:1996,Kitaev1995,Shor:1997}. For these problems, the time and memory resources needed to solve a given instance increase more slowly for the quantum algorithm than for the classical algorithm. Hence, irrespective of the actual wallclock time required for elementary operations on a quantum computer, there exists a problem size at which the quantum computer outperforms its classical counterpart. Discovery of new quantum algorithms remains challenging but the promise of known quantum algorithms has already motivated the development of many experimental approaches to the construction of a large scale quantum computer~\cite{Dykman2000,Knill2001,Haffner2008,Saffman2010,Mariantoni2011}.

Quantum simulation originated with Feynman's idea to use quantum devices to simulate quantum systems~\cite{Feynman1982}. Early work developed this idea through methods for simulating specific systems~\cite{Meyer1996,Wiesner1996,Lloyd1996,Lidar1997,Boghosian1998,Zalka1998} and by developing quantum algorithms for specific simulation tasks~\cite{Abrams1999,BACS07,Kassal2008,Wiebe2008,Ward2009,Sanders2012,Sanders2013}. Trapped ions, trapped atoms and photonic systems have recently been proposed for the simulation of quantum lattice models~\cite{Weimer2010,Ma2011,Hague2013,Cohen2013,Hauke2013}. These proposals have been realized experimentally for several systems~\cite{Simon2011,Greiner2009,wineland2002,Schaetz2008,Johanning2009,Ma2012b,Monroe2013}. 

Systems of interacting fermions are a natural target for quantum simulation. In particular, this implies the simulation of interacting electronic systems, due to the status of the electron as perhaps the most scientifically important fermion. In physics, a significant goal would be the quantum simulation of the phase diagram of the Fermi-Hubbard model due to its importance for high temperature superconductivity. In chemistry, the electronic structure problem provides a rich set of instances that can be addressed by quantum simulation~\cite{Aspuru-Guzik2006,Whitfield2010,Seeley2012,McClean2013,Kassal2010,CodyJones2012,Lanyon2009}.

Almost all methods to date for simulating systems of fermions and electronic structure take advantage of the second quantized formalism~\footnote{A notable exception to the use of the occupation number basis is the work of Whitfield using configuration state functions to perform spin-free simulation~\cite{Whitfield2013b}.
}. The overall Hamiltonian is expressed as a sum of combinations of creation and annihilation operators. The exchange symmetry of the problem is represented by the algebra of the creation and annihilation operators. The state of the system is represented using qubits in the occupation number basis, in which the state of each qubit represents the occupancy state of one orbital. The creation and annihilation operators are then mapped to qubit operators by, for example, the Jordan-Wigner or Bravyi-Kitaev transformations~\cite{Jordan1928,Somma2002,Seeley2012}. This approach to the simulation of electronic structure in particular has been widely explored~\cite{Aspuru-Guzik2006,Wang2008,Veis2010,Whitfield2010,Kassal2010,CodyJones2012,Yung2013}, and early examples have been implemented in NMR and optical quantum computers~\cite{Lanyon2009,Lu2011,McClean2013}. An adiabatic approach based on the same representation was also recently proposed~\cite{Babbush2013c}. 

The scaling of the number of qubits required for quantum simulation based on the second-quantized formalism is not optimal. For configurations of $n_e$ fermions that may occupy $n_o$ orbitals the second-quantized method of simulation requires $n_o$ qubits. However in chemistry the number of electrons is fixed for many problems of interest. In particular, for a chemical system interacting via the electronic structure Hamiltonian the number of electrons is a good quantum number, and so storage of all possible occupancies less than or equal to the number of molecular orbitals is unnecessary. In addition, the large number of terms $\simeq n_o^4$ in the interacting fermion Hamiltonian results in a large number of gates required for the simulation~\cite{CodyJones2012,Whitfield2010,Wecker2013}. This observation motivates the development of quantum algorithms for fermionic simulation and electronic structure whose qubit requirements scale with the number of electrons, and which open the door to improved scaling in the number of gates. This is the subject of the present paper.

Here, we investigate an alternative approach to the problem of simulating fermionic systems using the Full Configuration-Interaction (CI) matrix directly. The many electron wavefunction is expanded in a complete basis of Slater determinants corresponding to all possible configurations of $n_e$ electrons and $n_o$ orbitals. This basis is mapped to the qubits by indexing the Slater determinants and encoding the index into the computational basis of the qubits. The qubit requirements are therefore reduced from ${n_o}$ to $\lceil\log_2{n_o\choose n_e}\rceil$, which is optimal for the representation of all configurations of a fixed number of electrons. Time evolution in this basis is performed by exploiting the sparsity of the CI matrix.  By invoking the extensive methodology of quantum simulation of time evolution of sparse Hamiltonians it is possible to realize a number of different scalings of computational cost that trade time and space resources in ways that are not possible in the second quantized representation.

\section{The electronic structure problem}

The set of single-particle spatial wave-functions that comprise the molecular orbitals~$\left\{\phi_p (x) \right\}$, together with spin functions $\left\{\sigma_p(\omega)  \right\}$  form the set of spin orbitals $\left\{\chi_p(\bm{x}) \right\}$, where $\chi_p=\phi_p \sigma_p$ and where $\bm{x}=(x, \omega)$ denotes the pair of  spatial and spin parameters. The Hamiltonian in the Born-Oppenheimer approximation is comprised of one-electron and two-electron interaction operators. The one-electron operator for the $i^{\text{th}}$ electron takes the form 
 \begin{align}\label{estruc}
\hat h=\sum_{i=1}^{n_e}\hat h(i)=\sum_{i=1}^{n_e} \biggl[-\frac{1}{2}\nabla^2_i-\sum_A \frac{Z_A}{r_{iA}}\biggr], 
\end{align}
Thus, the Hamiltonian, in the appropriate units, takes the form:
\begin{align}
H=\sum_{i=1}^{n_e} h(i)+\sum_{i=1}^{n_e} \sum_{j> i}^{n_e}\frac{1}{r_{i,j}}. 
\end{align} 
The one-electron integrals are:
\begin{align}
h_{pq}(i)=\int \text{d}\bm{x}\:\: \chi^{*}_p(\bm{x})  \left(-\frac{1}{2} \nabla^2_i-\sum_A \frac{Z_A}{r_{iA}}\right)\chi_q(\bm{x}) ,   
\end{align}
and the two-electron integrals are: 
\begin{align}
h_{pqrs}(i,j)=\int \text{d}\bm{x}_1\text{d}\bm{x}_2\:\: \frac{\chi^{*}_p(\bm{x}_1) \chi^{*}_q(\bm{x}_2)\chi_r(\bm{x}_1) \chi_s(\bm{x}_2)}{r_{i,j}}.  
\end{align}

Given a set of orbitals, which are one-electron states obtained, for example, from a mean-field Hartree-Fock calculation, the next step is to construct a basis for the many electron states of the system. The Slater determinants form such a basis and are constructed to be anti-symmetric with respect to the interchange of any two electrons~\cite{SO96}. A Slater determinant containing two identical orbitals is zero. Each Slater determinant is specified by a {\em configuration} - an assignment of $n_e$ electrons to $n_e$ distinct orbitals chosen from the set of $n_o$ orbitals. The configuration labelling a Slater determinant is denoted as~$\ket{\chi_1, \cdots,\chi_{n_e}}$, where~$\chi_i$ denotes the orbital that the~$i^{\text{th}}$ electron occupies, for~$i \in \left\{1, \cdots, n_o\right\}$.  

The Hamiltonian~(\ref{estruc}) written in this basis is known as \emph{Configuration Interaction} (CI) matrix, and determination of the energy eigenvalues in the basis of all Slater determinants (equivalently all configurations) is known as the Full Configuration Interaction (FCI) method. We work with the CI-matrix representation of the Hamiltonian for the rest of the paper. The matrix elements of the CI matrix are determined by the Slater rules. The matrix elements between two configurations $x=\{\chi^1_i\}$ and  $y=\{\chi^2_i\}$ are given by:
\begin{enumerate}
\item{If $x=y$ we have a diagonal element given by
\begin{equation}
\bra{x}H\ket{x} = \sum_i^{n_e} \bra{\chi^1_i} \hat h\ket{\chi_i^1} +\sum_{i<j}^{n_e} \biggl( \braket{\chi^1_i\chi_j^1}{\chi^1_i\chi_j^1} - \braket{\chi^1_i\chi_j^1}{\chi^1_j\chi_i^1}   \biggr)
\end{equation}
}
\item{If $x$ and $y$ differ in one spin-orbital, $\chi^1_p$ in $x$ and $\chi^2_q$ in $y$
\begin{equation}
\bra{x}H\ket{y} =  \bra{\chi_p} \hat h\ket{\chi_q} +\sum_{l}^{n_e} \biggl( \braket{\chi^1_p\chi^1_l}{\chi^2_q\chi^2_l} - \braket{\chi^1_p\chi^1_l}{\chi^2_l\chi^2_q}   \biggr)
\end{equation}
}
\item{If $x$ and $y$ differ in two spin-orbitals, $\chi^1_p\chi^1_q$ in $x$ and $\chi^2_r\chi^2_s$ in $y$
\begin{equation}
\bra{x}H\ket{y} =   \braket{\chi^1_p\chi^1_q}{\chi^2_r\chi^2_s} - \braket{\chi^1_p\chi^1_q}{\chi^2_s\chi^2_r}  
\end{equation}
}
\item{If $x$ and $y$ differ in more than two spin-orbitals
\begin{equation}
\bra{x}H\ket{y} =  0
\end{equation}
}
\end{enumerate}

We now show that the CI-matrix is  sparse, {\em i.e.} most of its elements on each row and column are zero. In general, a $n\times n$ matrix is sparse if the number of non-zero elements on each row or column is a polylog function of $n$. The dimension of the CI matrix is $n_o\choose n_e$, and so if the number of nonzero entries is polynomial in $n_o$ and $n_e$ then the matrix is sparse. The total number of the non-zero entries in each row is given by~\footnote{The sparsity of the CI-matrix can be calculated using the Slater rules. For each given row, the diagonal element is non-zero. Next we consider all entries between orbital states that differ in only one orbital. The total number of such entries is equal to the number of ways that we can choose one of the $n_e$ occupied orbitals and one of the remaining $n_o-n_e$  unoccupied orbitals and interchanging them. Finally, we have to consider all entries between states that differ in two orbitals. The total number of the latter type  is equal to the number of ways that we can choose two of the $n_e$ occupied orbitals and two of the remaining $n_o-n_e$  unoccupied orbitals and interchanging them. Every other entry in the given row is zero because of the Slater rules.}:  
 \begin{align}
 \label{eq:d}
d =& {n_e\choose 2} {n_o-n_e\choose 2}+{n_e\choose 1} {n_o-n_e\choose 1}+1\\
   =& \frac{1}{4} n_e (n_e-1) (n_o-n_e)(n_o-n_e-1)+n_e (n_o-n_e)+1,  
\end{align}
which is quadratic in $n_o$ and quartic in $n_e$. The same is naturally true for each column as the Hamiltonian operator is Hermitian and hence the CI matrix is sparse.

\section{Sparse simulation algorithm}

The first systematic classification of Hamiltonians by their simulatability on quantum computers was given in~\cite{childsthesis}. Diagonal Hamiltonians, sparse Hamiltonians and Hamiltonians that can be efficiently put into diagonal or sparse form are considered there and in related works (see Table~\ref{tab1}). For a recent summary of progress in Hamiltonian simulation algorithms see~\cite{Sanders2013}. The first work on efficient simulation of sparse Hamiltonians whose non-zero  matrix elements are available through function evaluations was in~\cite{AT03}. Subsequently, a range of scalings of the number of qubits and gates required has been obtained by various authors. These are summarized in Table~\ref{tab1}.

\begin{table}[h]
\centering 
\begin{tabular}{|l|l|}
\hline
&\\
 Algorithm  & Number of evaluations  \\
 &\\
 \hline
  &\\
Aharonov and Ta-Shma~\cite{AT03}   & $O(n^9d^4 t^2\frac{1}{\epsilon})$  \\
 &\\
\hline
 &\\
Childs~\cite{childsthesis}   & $O(n^2d^{4+o(1)}t^{3/2}\frac{1}{\epsilon})$    \\
 &\\
\hline
 &\\
Berry, Ahokas, Cleve, Sanders~\cite{BACS07}&$O(\log^*nd^{4+o(1)}t^{1+2/k}\frac{1}{\epsilon^{1/2k}})$\\
 &\\
 \hline
&\\
Berry, Childs~\cite{Berry2012}&\\
For $\Lambda>||H||$, $\Lambda_{\rm max}>||H||_{\rm max}$&$O(\Lambda t \frac{1}{\epsilon^{1/2}}+d||\Lambda_{\rm max}||t)$ \\
If $\epsilon d>||H||t>\sqrt{\epsilon}$&$O(d^{2/3}[\log\log d||H|| t]^{4/3}\frac{1}{\epsilon})$ \\
If all terms have comparable norms&$O((\Lambda t)^{3/2}\sqrt{d}(\log d)^{7/4})\frac{1}{\sqrt{\epsilon}}$\\
&\\
\hline
&\\
Berry, Cleve, Somma~\cite{Berry2013a} &$O([d^2 ||H||t + \log\frac{1}{\epsilon}]\log^3[d(||H||+||H'||)]n^c)$\\
&for c a constant\\
&\\
\hline
&\\
Berry, Childs, Cleve, Kothari, Somma~\cite{Berry2013b}&$O(\tau\frac{\log(\tau/\epsilon)}{\log\log(\tau/\epsilon)})$ and $O(\tau\frac{\log^2(\tau/\epsilon)}{\log\log(\tau/\epsilon)n})$ two qubit gates
$\tau=d^2 ||H||_{\rm max}$\\
&\\
\hline
\end{tabular}
  \caption{Summary of various approaches and their associated scaling with sparsity, time and error. For the CI matrix the sparsity is given by $d=(1/4) n_e (n_e-1) (n_o-n_e)(n_o-n_e-1)+n_e (n_o-n_e)+1$.}\label{tab1}
\end{table}

In the limit of high accuracy for a fixed number of electrons, where $n_o\gg n_e$, the sparsity is quadratic in $n_o-n_e$. Use of the methods of~\cite{BACS07} results in scaling that is $(n_o-n_e)^8$ - worse than that obtained from the second quantized approaches. Use of the methods of~\cite{Berry2012} in the general case can give a scaling of $(n_o-n_e)^{4/3}[\log\log (n_o-n_e)]^{4/3}$, but in the case where all terms in the decomposition have comparable norms one can realize an almost linear scaling of $(n_o-n_e)[\log (n_o-n_e)]^{7/4}$. This decrease in the number of terms in the Trotter expansion is significant from an asymptotic point of view, given that the occupation number methods inevitably result in a number of terms scaling as $n_o^4$. The methods proposed here do impose some additional fixed overhead required to represent the Hamiltonian matrix elements in additional qubits, proportional to the number of bits of precision required to represent the Hamiltonian matrix elements and to the logarithm of the sparsity. For quantum chemical problems of interest these additional qubits could require a few tens of extra qubits.  

This scaling with fixed number of elecrons and growing number of orbitals, the limit of high accuracy for a given molecule, should be contrasted with scaling at constant filling fraction. In this case $n_e$ and $n_o$ are proportional. The sparsity is then quartic in $n_e$ (equivalently in $n_o$), and the corresponding scalings follow from Table~\ref{tab1}, as discussed in~\cite{Wecker2013}.There are therefore two scaling limits that are conceptually different. The limit of fixed $n_e$ and growing $n_o$ is the continuum limit for a molecule of fixed size - the limit where the deviation of the CI energy from the true value due to finite basis set size vanishes. The limit of where $n_o$ grows together with $n_e$, where the filling fraction is held constant, is conceptually related to the infinite system size or thermodynamic limit where one seeks bulk properties at fixed accuracy. Here the error due to the finite extent of the modelled system is going to zero. Here we are invoking the continuum limit scaling, meaning we are considering the case of high accuracy calculations of a particular molecule. The quartic scaling in $n_e$ suggests that the methods we describe in the next section will be appropriate for small molecules.

\subsection{Outline}
\label{subsec:Outline}
 
Our aim is to simulate a system comprised of many electrons based on the Configuration-Interaction (CI) formalism, taking advantage of the sparsity of the CI matrix to reduce the number of qubit resource required.  Following~\cite{AT03}, we say the Hamiltonian H on~$n$ qubits is simulatable if  the unitary $U(t)=e^{-\imath H t}$ for every time~$t>0$, can be approximated within any given error $\epsilon$, where $0<\varepsilon<1$, by a quantum circuit of size $\text{poly}(n, t, 1/\varepsilon)$. We first devise an efficient gate-model circuit  using the results of  Suzuki~\cite{S93},  Aharonov and Ta-Shma~\cite{AT03} and Berry {\it et. al}~\cite{BACS07}. This algorithm requires at most $d^2$ colors for the decomposition of the Hamiltonian, resulting in at most $d^2$ one-sparse terms. We then refine this algorithm using properties of the CI matrix in order to achieve a decomposition with $d$ terms. 

The algorithm involves the following steps: 
\begin{itemize}
\item Finding a coloring scheme for the edges of the graph associated with the CI-Hamiltonian such that no two edges incident on the same node have the same color. We prove that this is property of the CI matrix implied by its definition via the Slater rules.

\item Decomposing the Hamiltonian into a sum of one-sparse Hamiltonian operators~$H=\sum_m H_m$, each labelled by a graph-color $m$. 

\item Using the higher-order Trotter-Suzuki-decomposition method to approximate the evolution of the overall Hamiltonian as a product of the non-commuting unitaries~$U_m=e^{-\imath H_m t}$. 
 
\item Converting each one-sparse Hamiltonian into a series of gates that simulate the unitary evolution $U_m=e^{-\imath H_m t}$ associated with the Hamiltonian~$H_m$. 
 
\end{itemize}

\subsection{The qubit representation of the CI-matrix}
\label{subsec:Qubit}

There are two representations of fermionic states that are widely used. The first is the occupation number basis, first introduced in the context of quantum information in~\cite{Zanardi2002} and used for quantum simualtion of quantum chemical systems in~\cite{Somma2002,Aspuru-Guzik2006}. In this representation one qubit is assigned per orbital, and the occupation of the orbital is denoted by the state of the qubit, $1$ for occupied and $0$ for unoccupied. Each possible occupancy basis state of the full Fock space can therefore be labelled by the integer whose binary expansion gives the occupation state of the orbitals:
\begin{align}
\ket{k} = \ket{k_{n_o}...k_{1}} = \ket{\sum_{i=1}^{n_o}k_i2^{i-1}}.
\end{align}
This provides a simple and direct mapping from the logical basis states of the qubits to the occupancy states of the orbitals. There are two drawbacks to this mapping. Firstly, in order to correctly represent a second quantized Hamiltonian one must define qubit creation and annihilation operators that obey the same anticommutation relations as the fermionic operators. This requires the use of the Jordan-Wigner or Bravyi-Kitaev transformation that impose some overhead in the number of quantum gates required to simulate time evolution under the Hamiltonian. Secondly, the entire Fock space is mapped to the Hilbert space of the qubits, so the qubit requirements scale with the number of orbitals. However, in many applications of interest the Hamiltonian commutes with the number operator and so the number of particles is a good quantum number. For example, in the electronic structure problem in quantum chemistry one is typically interested in a problem for a fixed number of electrons. Hence there is considerable redundancy in qubit requirements in the use of the occupation number representation. 

The second representation labels each Slater determinant by the list of occupied orbitals. Numbering the orbitals by integers from $1$ to $n_o$ the basis states for a fixed number of $n_e$ electrons are labelled by the $n_e$ element subsets of $L=\{j|1\leq j\leq n_o\}$. 
A label can be assigned to each basis state as follows:
\begin{align}
\ket{\chi_{n_e},\dots,\chi_{1}}  \equiv \ket{\sum_{i=1}^{n_e}2^{\chi_i-1}}=\ket{x} ~~~\chi_i\in L~\forall~ i
\end{align}
in this case the labels 
\begin{align}
\label{eq:xchi}
x=\sum_{i=1}^{n_e}2^{\chi_i-1}
\end{align}
are non-consecutive integers each of which has $n_e$ ones in their binary expansion. There are:
\begin{align}
D(n_o,n_e)={n_o \choose n_e}
\end{align}
of these states, this being the dimension of the $n_e$ particle subspace of the Fock space. We can map this space to a subspace of the Hilbert space of $n_Q= \lceil \log_2 D(n_o,n_e) \rceil$ qubits by assigning the first $D(n_o,n_e)$ logical qubit states to the states $\ket{x}$ in order.  Let us denote the index of the state $\ket{x}$ in this ordering scheme by $q(x)$. Conversely, we associate the state label associated with the number $q$ in the list with $\ket{x(q)}$. Note that the map $q: x\mapsto q(x)$ is a   bijection.  Thus, we are dealing with a two fold encoding of each Slater determinant basis state: 
\begin{align}
\ket{\chi_{n_e},\dots,\chi_{1}} \rightarrow \ket{x}\rightarrow \ket{q(x)}
\end{align}
and decoding is similarly twofold:
\begin{align}
\ket{q}\rightarrow \ket{x(q)}\rightarrow \ket{\text{orb}(x)},
\end{align}
where we denote by 
\begin{align}
\label{def:spec}
\text{orb}(x):=\left\{\chi_1, \dots, \chi_{n_e} \right\}
\end{align}
 the configuration of  the state $\ket{x}$, {\it i.e} the set containing  the particular $n_e$ orbitals in the Slater determinant that is associated with the state $\ket{x}$, and, by extension, the label $q(x)$. 
This mapping is optimal in terms of the number of qubits required for storage of the $n_e$ electron wavefunction.

\subsection{Decomposing the CI-matrix}
\label{sec:Deccomp}

Given a mapping of the basis of Slater determinants to the logical basis of our qubits the next question is whether the time evolution under the CI-matrix can be simulated efficiently by a quantum algorithm. In the case of the occupation number representation this question has been answered in the affirmative, giving simulation with qubit requirements at least $n_o$ and a number of elementary gates scaling polynomially~\cite{Jordan1928,Somma2002,Seeley2012}. The occupation number representation admits a natural tensor product structure - each orbital is mapped to a single qubit, which is a single tensor factor of the Hilbert space of our computer. The Hamiltonian is decomposed into a sum of $O(n_o^4)$ terms each of which act on $O(n_o)$ or $O(\log n_o)$ qubits, and in turn the evolution operator may be decomposed by Suzuki-Trotter techniques. The tensor product structure also gives us the usual definition of ``elementary'' operations: quantum gates which act on one or two qubits at a time~\cite{MikeIke}. In spite of this natural decomposition the overheads arising from Trotterization may be significant. Recent numerical work aimed at empirically estimating the gate cost in detail for a full phase estimation algorithm reaching chemical accuracy indicate that the occupation-number based algorithms require extremely deep circuits~\cite{Wecker2013}.

In the case of the CI-matrix, simulation of the time evolution is more challenging from the point of view of algorithm design. The basis of Slater states has no natural tensor product structure, and the CI-matrix is defined through the Slater rules, and not through a sum over a polynomial number of terms. However, as we pointed out earlier, the CI-matrix is $d$-sparse where  $d$ is given in~Eq. (\ref{eq:d}). The locations of the nonzero matrix elements are straightforward to determine, and their calculation is efficient. Hence simulation of the CI-matrix is a good target for techniques developed for the quantum simulation of sparse matrices~\cite{AT03, BACS07, WBHS11}.

In the absence of a tensor product structure one must define what the elementary operations are - what are the unitary operators into which one will decompose the full time evolution operator? In the case of sparse Hamiltonians the elementary operations are one-sparse - that is, they have at most one non-zero element in each row and each column~\cite{AT03}. Such matrices can be implemented efficiently given an oracle that generates their non-zero entries. This is entirely consistent with the definition of elementary operations in the case where there is tensor product structure to the Hamiltonian. For example, any tensor product of Pauli matrices is one-local, and hence local tensor product decompositions are a special case of more general sparse decompositions.

The first step in the simulation of a sparse Hamiltonian is therefore to decompose the Hamiltonian into one-sparse terms. We now discuss the methods for decomposing the CI-Hamiltonian matrix $H$ into one-sparse matrices $H=\sum_m H_m$  since the evolution of each $H_m$ can be separately simulated efficiently.  

\subsection{The standard graph-coloring method}

We represent the Slater determinants by nodes in a graph, where two nodes are connected if the matrix element between the corresponding determinants is not zero. Each matrix element $\bra{a}H\ket{b}$ for $a>b$ corresponds to an undirected edge in the graph. For the case of Hermitian matrices the graph is undirected because $\bra{a}H\ket{b}=\bra{b}H\ket{a}^*$. A decomposition of a sparse Hamiltonian into a sum of one-sparse Hamiltonians is therefore a division of the graph representing the Hamiltonian into a set of disjoint subgraphs. This is equivalent to a coloring of the edges of the graph, such that distinct colors label subgraphs corresponding to one-sparse Hamiltonians. Here, we present a a graph coloring based on that used in~\cite{AT03, BACS07, WBHS11}. What is the graph corresponding to a one-sparse Hermitian matrix? Each row and column of the matrix has only one nonzero element.  The nodes are divided into a collection of disjoint pairs $(a,b)$ connected by the edges corresponding to $\bra{a}H\ket{b}=\bra{b}H\ket{a}^*$. The degree of the graph corresponding to a one sparse matrix is one. 

As described in Section~\ref{subsec:Qubit} and in Appendix~\ref{sec:Agraph}, we denote each computational basis state associated with the Slater determinant $\ket{x}$ with $\ket{q(x)}$. Also, let us denote the node in the graph corresponding to $\ket{x}$ by $q(x)$. We identify each color label with an ordered pair, $\bm{k}=(i,j)$, as follows. First construct the {\em directed} graph in which we distinguish between incoming and outgoing edges. We assume each node has $d$ edges, and the outgoing edges from each node are numbered from $1$ to $d$ according to the order of the labels of the nodes they are connected to. This assigns a label between $1$ and $d$ to each directed edge in the graph. We assign the pair of edge labels $(i,j)$ to the undirected edge in the corresponding undirected graph. Hence nodes $x$ and $y$ are connected by an edge $(i,j)$ if $y$ is $x$'s $i$th neighbor and $x$ is $y$'s $j$th neighbor.

We next form a one-sparse matrix $H_{m(\bm{k})}$ associated with each color, where $m$ is  the index that counts all possible color labels $\bm{k}$. The  range of $m$  is $\left\{1 \ldots d^2\right\}$. In particular, we assign  
\begin{align}
\begin{cases}
H_m\left(q(x_i),q(y_j)\right):=\bra{x_i} H\ket{y_j}\\
H_m(q(y_j),q(x_i)):=\bra{x_i} H^* \ket{y_j}=\bra{x_i} H\ket{y_j}, \;\;\; \text{(The orbitals can always be chosen to be real)},   \\
H_m(q(x_i),q(y')):=0, \: y' \neq y_j\\\:\:  
H_m(q(y_j),q(x')):=0, \: x' \neq x_i.\:\:  
 \end{cases}
\end{align}  

For a general Hamiltonian a third index in the color label is needed, as the pair $\bm{k}=(i,j)$ determined by the procedure could result in two adjacent edges having the same color pair assigned to them~\cite{BACS07}. This happens when three nodes exist, say $a, b, c$, where $a \leq b\leq c$, such that  $b=a_i$ and $a=b_j$ and, at the same time, $c=b_i$ and $b=c_j$, for the same integers $i$ and $j$.  Therefore, the general color label is a triple $\bm{k}=(i,j;v)$ where the third index $v$ is added to differentiate between the labels of adjacent labels in such cases. However, in the case of the CI-matrix because of the conditions set by the Slater rules, we need not add a third label because, the graphs associated with CI-matrices will never have two adjacent nodes with the same color label following the method presented here (See Appendix~\ref{sec:Agraph} for a proof). Note that we have categorized  labels that connect nodes that differ in a single orbital in their Slater representation as one type of label from the outset.  Labels that connect nodes  differing in two orbitals are likewise marked separated beforehand. Hence,  we no longer need to check for coincidences in labels from one category and the other in this scheme.  

\subsection{The band-coloring method}

In general the method above requires at most $d^2$ colors for the graph, and hence decomposes the Hamiltonian into a sum of at most that many terms. However, it is clearly possible that not all of the $d^2$ colors occur in the labelling of a given graph. For graphs arising from the CI matrix, we can show that in fact only $d$ colors are actually required. Once again, this is due to the structure of the graph arising from the Slater rules that define the CI matrix.

Nodes are again labelled by orbital configurations and edges are present if two orbital configurations differ by only one or two orbitals. That is, two nodes are connected by an edge if they are related by application of one or two  ``excitation operators"~\footnote{in this context, as these operators act on an arbitrary configuration, they may in fact lower the energy of a configuration.}  that deletes one orbital in the configuration and replaces it by another. 
\begin{align}
\{\chi_i\} \rightarrow \{\chi_i\}\setminus \chi_{u} \bigcup \{\chi_d\}. 
\end{align}

If the node labels of a pair of vertices are $q(x)$ and $q(y)$, where $q(y)$ is reached from $q(x)$ by exchanging $\chi_d$ with $\chi_u$, what is the condition on $\chi_u$ and $\chi_d$ when $q(y)>q(x)$? Since $q(x)$ and $q(y)$ are indices of $\ket{x}$ and $\ket{y}$ labels in increasing order, the condition $q(y)>q(x)$ implies $y>x$, the map $q:x\mapsto q(x)$ being a monotonically increasing map.   If we consider the occupied orbital representation of the configurations in Eq.~(\ref{eq:xchi}), it is clear that the action of excitation operators is to exchange the term $2^{\chi_d-1}$ in $x$ with $2^{\chi_u-1}$ in $y$. Given this, it is evident that:
\begin{align}
y>x~~{\rm iff}~~\chi_u>\chi_d.
\end{align}

 Finally, for a given node $q(x)$ and a given orbital index $i \in [1,n_o]$,  we define
 \begin{align}
 O^{\text{incl}}(x)&:=\left\{j; \:\:1\le j\le n_e, \:\chi_j \in \text{orb}(x)\right\},\\
 O^{\text{excl}}(x)&:=\left\{j; \:\:1\le j<\le n_e \:\chi_j \notin \text{orb}(x)\right\}, 
 \end{align}
 to denote the set containing the orbitals below the orbital $i$ that are included in the configuration of $\ket{x}$ (see (\ref{def:spec})), and the set of orbitals below orbital $i$ that are excluded from the configuration of $\ket{x}$, respectively. Note that there are $n_e$ elements in $O^{\text{incl}}(x)$ and $n_o-n_e$ elements in $ O^{\text{excl}}(x)$.  At this stage we wish to introduce a numerical edge label. 
The edges of the graph representing the Hamiltonian are labelled according to a numbering scheme as follows:

\begin{enumerate}
\item{For all edges of the form~$x\rightarrow x$ we assign the label 0:
\begin{align}
e_{x \rightarrow x}=0, \:\: \forall x. 
\end{align}
}
\item{If~$q(x)$ and~$q(y)$ are two nodes that differ in the position of one occupied orbital, then the edge between them is labeled as 
\begin{align}
e_{x \rightarrow y}=(d,u),  
\end{align}
where the~$d^{\text{th}}$ occupied orbital in~$\ket{x}$, i.e.~$\chi_d \in \text{orb}(x)$,  is unoccupied in $\ket{y}$,  so that $d \in  O^{\text{incl}}(x)$ and $d \in O^{\text{excl}}(y)$, and instead the~$u^{\text{th}}$ occupied orbital in $\ket{y}$, i.e.~$\chi_u \in \text{orb}(y)$ is unoccupied in $\ket{x}$ and $u \in  O^{\text{incl}}(y)$ and $u \in O^{\text{excl}}(x)$. 
There are $n_e$ orbitals occupied in either $\ket{x}$ or $\ket{y}$, and there are $n_o-n_e$ non-occupied orbitals in either one. So,the number of exchanges between the two that is the total number of edge labels of this kind is $n_e (n_0-n_e)$. The colors in this class are labelled by $m=1,\dots,n_e(n_o-n_e)$.
 } 
\item{Finally, if~$p(x)$ and~$p(y)$ are two nodes that differ in two occupied orbitals, then the edge between them is labeled as 
\begin{align}
e_{x \rightarrow y}=(d_1,u_1; d_2, u_2 ), 
\end{align}
where the $d_1^{\text{th}}$ and $d_2^{\text{th}}$ orbitals  in $\text{orb}(x)$ are replaced with the $u_1^{\text{th}}$ and $u_2^{\text{th}}$  orbitals. Similar to the previous type of edges, the total number of edges of this third type is equal to the number of ways of swapping two orbitals in $O^{\text{incl}}(x)$ and two orbitals in $O^{\text{excl}}(x)$ to transform $\ket{x}$ to $\ket{y}$.  
The total number of this type of edge labels is therefore equal to ${n_e \choose 2} {n_o -n_e\choose 2}$. We label each color in this class by $m=n_e(n_o-n_e)+1,\dots,n_e(n_o-n_e)+{n_e\choose 2}{n_o-n_e\choose 2}$.}
\end{enumerate}
Thus, the total number of edge labels we need in this scheme is equal to the sparsity of the CI-matrix. 
\begin{align}
d={n_e \choose 2} {n_o -n_e\choose 2} + n_e (n_o-n_e) +1,   
\end{align}
In other words, taking advantage of the very specific structure of the CI-matrix allows us to have a labelling scheme that is linear in the sparsity $d$. 
 
\subsection{Trotterization}
\label{sec:Trotter}
The coloring scheme defined above defines a decomposition of the Hamiltonian $H=\sum_m H_m$ into simpler summands and we can use the Trotter method to write the time evolution as a product of evolutions under the individual terms $H_m$. The time evolution of the total Hamiltonian between time  $t_0$ and $t_0+\Delta t$  can be expressed as 
\begin{align}
U(t_0, t_0+\Delta t) =\tau \exp\left(-\imath \int _{t_0}^{t_0+\Delta t} H(u) du\right)
\end{align}
The time evolution $U(t_0, t_0+\Delta t)$ can then be approximated in terms of $U_m(\Delta t):=\exp(-\imath H_m\Delta t/2) $. 
The first order approximation is 
\begin{align}
U^{(1)}(t_0, t_0+\Delta t)=\prod_{j=1}^{m_{\max}} \exp(-\imath H_m\Delta t/2) =\prod_{j=1}^{m_{\max}} \exp(-\imath H_m\Delta t/2). 
\end{align}
Higher order approximations are derived recursively from lower order approximations as 
\begin{align}
U^{(\ell)}(t_0, t_0+\Delta t)=U^{(\ell-1)}(t_0+(1-s_\ell)\Delta t, t_0+\Delta t) U^{(\ell-1)}(t_0+(1-2s_\ell)\Delta t, t_0+(1-s_\ell)\Delta t) \\
\times U^{(\ell-1)}(t_0+s_\ell\Delta t, t_0+(1-2s_\ell)\Delta t) U^{(\ell-1)}(t_0+s_\ell\Delta t, t_0+2 s_\ell\Delta t) U^{(\ell-1)}(t_0,t_0+s_\ell\Delta t),
\end{align}
where the value of $s_\ell$ are given by $s=1/(4-4^{1/(2\ell-1)})$~\cite{S93}. The Trotterization procedure employed in performing the time steps is independent from the details of the decomposition, and every new enhancement of the procedure can be immediately employed in the scheme. We have seen so far how time evolution under  $H=\sum_m H_m$ can be expressed in terms of evolution under  each individual one-sparse Hamiltonian $H_m$. Next we consider how to efficiently simulate time evolution under one-sparse Hamiltonians, first considering the oracles which define the non-zero matrix elements.

\subsection{Evaluating the matrix elements}
\label{sec:Oracle}

For each one-sparse matrix in the decomposition of the CI-matrix $H=\sum_m H_m$  we define a function that outputs the non-zero elements in the corresponding $H_m$ one bit at a time. We construct this from two functions, one that identifies the elements of the desired $H_m$, and one that returns their value. We follow the method outlined by Wiebe {\it et al.} in~\cite{WBHS11}.  The first function $\left\{Col_m \right\}$ outputs the column number of the non-zero element in row $q(x)$ of $H$ that appears in $H_{m}$ in bit wise manner. Each function $Col_m (x,p)$ returns the $p^{\text{th}}$ bit of the binary code for node $q(y)$ that is connected to $q(x)$ by an edge with color label $m$. It is defined by:
\begin{align}
QCol_{m} (p) \ket{q(x)}\ket{0}^{\otimes n_Q}=\ket{q(x)}\ket{Col_{m} (x,p)}, 
\end{align}
Where $n_Q$ is the size of the register that encodes the distinct colors required to label the graph, which is $d$ using the band scheme described above. Hence $n_Q=O(\lceil\log d\rceil)$. The second type of oracle is the function $Val(x,y,p)$ that returns the $p^{\text{th}}$ bit of the binary encoding of the CI-matrix element $H_{x,y}:=\bra{x}H\ket{y}$. Every element $\bra{x}H\ket{y}$ of the CI-matrix is real valued, and we use the polar representation of $H_{x,y}$ following~\cite{WBHS11}. Let $H_{x,y}=\exp(\imath \pi s_{x,y}) |H_{x,y}|$, where $|H_{x,y}|$ is the absolute value of of $H_{x,y}$ and  $s_{x,y}=0$ if $H_{x,y}$ is positive and $s_{x,y}=1$ if $H_{x,y}$ is negative. We encode $H_{x,y}$  using $n_H$ bits, with $1$ bit encoding $s_{x,y}$ and $n_H-1$ bits encoding $|H_{x,y}|/H_{\max}$, where $H_{\max}=||H||_{\max}$ is the largest element of $H$, also known as the maximum norm. The action is:
\begin{align}
QVal(p) \ket{q(x)}\ket{q(y)}\ket{0}^{\otimes n_H}=\ket{q(x)}\ket{q(y)}\ket{Val(x,y,p)}. 
\end{align}

The functions $Val(x,y,p)$ are the bit values of the CI-matrix elements given by the Slater rules. These values can be computed by a classical reversible circuit which would be included in the coherent quantum circuit realizing the full algorithm. We leave the details of this for future work, but note that this also raises the possibility of the computation of the integrals themselves within the quantum computation so that the input to the algorithm would be basis set parameters for $n_o$ orbitals rather than $n_o^4$ integrals, albrit with additional overhead. Likewise, computing the functions $Col_{m}$ is classical and efficient and can be broken down into reversible classical gates, and therefore efficiently implemented within a coherent quantum circuit. We require an additional $n_Q=2\log d$ and $n_H$ qubits to store the color labels and matrix element values. The additional qubit requirement for storing the colors grows as $n_Q=\log d$ using a coloring scheme with at most $d$ colors, and the precision requirements for the matrix elements are set by the requirement of precision for the overall problem. For example, for chemical problems the precision requirement is often taken to be fixed by ``chemical accuracy'' of $1$kCal per mol. This imposes a significant, but fixed overhead of a few tens of qubits to represent the matrix elements to sufficient precision. We note that it is the additional overhead in these parts of the circuit that will determine in practice where these algorithms offer practical advantages over other techniques.

\begin{figure}[h] 
\includegraphics{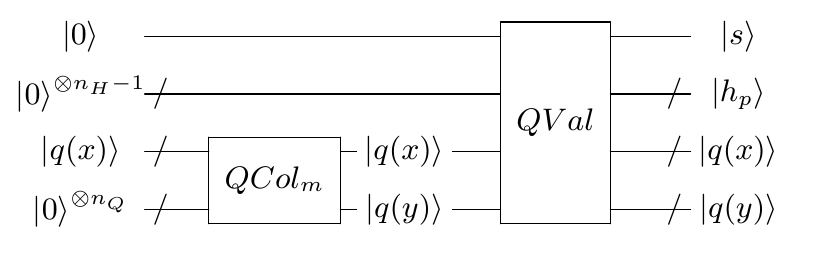}
 \caption{The circuit for uploading the values for the non-zero elements of the one-sparse Hamiltonian $H_m$ via two series of oracle calls $QCol_m=\left\{QCol_m(p)\right\}$ and $QVal=\left\{QVal(p)\right\}$.}
 \label{fig:fig0} 
\end{figure}

\subsection{Simulating One-Sparse Hamiltonians}
\label{sec:1-sparse}
In this section, we discuss the details of simulating the one-sparse Hamiltonians that appear in the decomposition of the CI-matrix. We assume, as before, that we have access to elements of each $H_m$ in a bit-wise manner, via successive calls to quantum oracles $\left\{QCol_{m}\right\}$ and $QVal$ that act on the computational basis which is, in turn, the binary encoding of the Slater states. In the next section, we go over the methods of combining the simulations of one-sparse Hamiltonians to simulate the time evolution of the overall complete Hamiltonian. We separate two cases. First, we consider the terms in the Hamiltonian $H_{\bm{0}}$ comprised of the  diagonal elements of $H$. Next, we consider the remaining one-sparse Hamiltonians in the decomposition of $H$.

\subsubsection{Diagonal elements of the CI-matrix}

Quantum simulation of diagonal terms of a Hamiltonian is a much-studied problem in the context of first-quantized or grid-based quantum simulation algorithms~\cite{Wiesner1996,Boghosian1998,Meyer1996,Kassal2008}. In these algorithms the multiparticle wavefunction is represented on a grid of points and the Hamiltonian is composed of kinetic and potential energy terms. The potential energy is diagonal in the position basis and the kinetic energy is diagonal in the momentum basis. One may transform between these bases efficiently using the quantum fourier transform, and so the entire simulation problem reduces to one of simulating diagonal Hamiltonians. These may be simulated efficiently by phase-kickback techniques, as described in~\cite{Kassal2008}. The simulatability of diagonal Hamiltonians in the more general setting relevant here was first noted in~\cite{childsthesis}.

\subsubsection{Off-Diagonal elements of the CI-Matrix}

\begin{figure}[center] 
\includegraphics{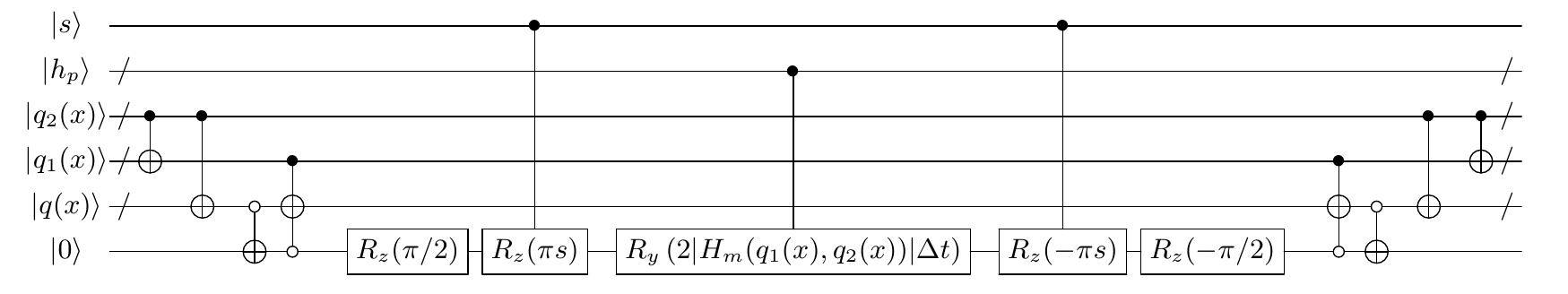}
\caption{The circuit for simulating the time evolution $\text{exp}(\imath H_{m} \Delta t)$ of a one-sparse Hamiltonian $H_m$. The values $s$ and $h_p$ are the bits specifying the sign and magnitude of $H_m(q_1(x),q_2(x))$.}
\label{fig:fig1} 
\end{figure}

The simulation method starts by taking an initial state $\ket{\psi}$ and, following closely the outline of the method in~\cite{WBHS11},  mapping it  by calling the oracle $QCol_{m}$,  
 to the state, 
\begin{align}
\ket{\psi}=\sum_x c_x \ket{q(x)}\ket{0}^{\otimes 2n_Q}\mapsto \sum_x c_x \ket{q(x)} \ket{q_1(x)} \ket{q_2(x)},
\end{align}  
where $x$ is a row index of the CI-matrix. If we denote the column number associated to $x$ in the coloring scheme labeled by $m$ by $y$, then 
\begin{align}
q_1(x)=\min(q(x),q(y)), \:\:\: q_2(x)=\max(q(x),q(y)). 
\end{align}
Next, we correlate the one or two-dimensional subspace spanned by $\left\{\ket{q_1(x)}, \ket{q_2(x)}\right\}$ to an ancillary qubit that is then evolved according the matrix element $H_{x,y}$. The outcome is equivalent to evolving the entire subspace by the one-sparse Hamiltonian $H_{m)}$.  We do this via the mapping 
\begin{align}
\sum_x c_x \ket{q_1(x)}\ket{q_2(x)} \ket{0} \mapsto \sum_x c_x \ket{q_1(x) \oplus q_2(x)} \ket{q_2(x)} \ket{a_x}, 
\end{align} 
where 
\begin{align}
a_x=
\begin{cases}
1, \:\:\: q_2(x)=q(x)\\
0, \;\;\; q_1(x)=q(x). 
\end{cases}
\end{align}
The final bit, $\ket{a_x}$ is the target qubit that is now correlated with the states $\ket{q(x)}$ in $\ket{\psi}$. In the next step, the qubit $\ket{a_x}$ is acted on by the elementary gates. The overall circuit is depicted in Figure~\ref{fig:fig1}. 

The evolution operator in this case is a rotation and  can be expanded as 
\begin{align}
\label{Uk2}
U_{m} (\Delta t) =\exp
\left(-\imath 
\begin{bmatrix}
0&H_m (q_1(x), q_2(x))\\
H^{*}_m (q_1(x),q_2(x))&0
\end{bmatrix}
\Delta t 
 \right). 
\end{align}
where $s \in \{0,1\}$ and the absolute value $|H_m(q_1(x), q_2(x))|$ specify the real Hamiltonian element $H_m(q_1(x), q_2(x))=\exp(\imath \pi s)  |H_m(q_1(x), q_2(x)|$ that is obtained bit by bit by applying the consecutive $QVal$ gates.  The time evolution $U_{m} (\Delta t)$ can then be expanded in terms of rotation operators as 
\begin{align}
\label{Uk2'}
U_{m} (\Delta t) = R_z(-\pi/2) R_z(-\pi s) R_y(2|H_m(q_1(x), q_2(x))| \Delta t) R_z(\pi s) R_z(\pi/2).   
\end{align}
As before, we can expand the above rotation as a sequence of Pauli operators,  and in particular, in terms of X and Z rotations. The  rotations can be implemented by a successive rotations of fixed degrees controlled by the qubits encoding the bits of $H_m(q_1(x),q_2(x))$. 

\section{Conclusions}
\label{sec:Con}
In this paper, we devised a quantum algorithm that simulates a system of fermions with Coulomb interactions using the CI-matrix representation and the Slater rules in quantum chemistry directly. Unlike previous methods of based on the second-quantized formalism we were able to reduce the number of qubits by storing only the $n_e$ particle sector of the full Fock space. We obtained algorithms that require the minimal number of qubits to represent the wavefunction while retaining the polynomial scaling of the number of gates in the quantum circuit.

The CI matrix is decomposed into a sum of one-sparse matrices using a coloring scheme that exploits the particular structure of the associated graph imposed by the Slater rules that define the CI matrix. We were able to show two coloring schemes that exploit the structure of this graph. The first scheme uses $d^2$ labels (an improvement over the generic case), while the second reduces this to $d$, for sparsity $d$. This is the first time, to our knowledge, that the graph structure associated with the Slater rules is studied. We believe that a detailed study of this structure in the future can be of interest to quantum chemists even for conventional computational approaches to quantum chemistry.

The purpose of the present paper is to obtain minimal qubit methods and investigate the efficacy of sparse Hamiltonian techniques for quantum simulation of electronic structure based on the CI-matrix. We obtained results relevant to the asymptotic scaling of the methods, and we show how the structure of the CI matrix can be used to obtain improved scaling in sparsity over the generic case. The determination of the prefactors and actual number of gates require numerical analysis of these algorithms for a suite of particular cases of interest, as was performed recently for the occupation number, second quantized methods in~\cite{Wecker2013}. The results obtained here for asymptotic scaling give some grounds for optimism, however, particularly when one considers the case of large numbers of orbitals for a fixed number of electrons. This is the appropriate limit for achieving high accuracy for a specific molecule. The qubit requirements to represent the wavefunction are optimal, the number of terms in the Trotterization for fixed number of electrons and increasing number of orbitals scales as $(n_o-n_e)^{4/3}\log\log (n_o-n_e)$ in the general case, and is at best (assuming equal norm terms) $(n_o-n_e)[\log (n_o-n_e)]^{7/4}$ using the black-box algorithms with the best scaling with sparsity developed in~\cite{Berry2012}.  Given the importance of precision to quantum chemical calculations it remains to be seen whether the methods offering exponentially improved precision are in fact superior in spite of their inferior scaling with sparsity~\cite{Berry2013a,Berry2013b}. It also remains to be seen whether these methods~\cite{Berry2013a,Berry2013b} can be systematically improved to offer better scaling with sparsity, as was the case in~\cite{AT03,BACS07,Berry2012} One sees that it is therefore possible to simultaneously reduce the number of  qubits required to represent the wavefunction and also change the scaling of the number of gates as compared with the methods proposed in~\cite{Aspuru-Guzik2006} and studied in~\cite{Wecker2013}.

\section*{Acknowledgements}

PJL and BT would like to thank Al\'an Aspuru-Guzik for productive discussions and for his hospitality during a visit to Harvard during which part of this work was completed. BT would like to thank Nathan Wiebe for helpful discussions. This project is supported by NSF CCI center, ÒQuantum Information for Quantum Chemistry (QIQC)Ó, award number CHE-1037992, by NSF award PHY-0955518 and by AFOSR award no FA9550-12-1-0046.

\appendix

\section{The graph-coloring method}

\label{sec:Agraph}

The edges of the graph representing the Hamiltonian are labelled according to a numbering scheme as follows. We label the edge connecting $q(x)$ to $q(y)$, where $x<y$, as an ordered pair $\hat{e}_{x,y}:=(e_{x\rightarrow y}, e_{y\rightarrow x})$. The pair of edge labels are defined as: 
\begin{align}
e_{y\rightarrow x}~~{\rm number~of~nodes} ~x'\leq x~{\rm to~which}~y~{\rm is~connected,}
\end{align}
and:
\begin{align}
e_{x\rightarrow y}~~{\rm number~of~nodes} ~y'\leq y~{\rm to~which}~x~{\rm is~connected.}
\end{align}
In addition, each node is connected to itself, with the edge label $\hat{e}_{x,x}:=(e_{x\rightarrow x}, e_{x\rightarrow x})$, where similarly, 
\begin{align}
e_{x\rightarrow x}~~{\rm number~of~nodes} ~x'\leq x~{\rm to~which}~x~{\rm is~connected.}
\end{align}
We now proceed to determining the formulas for direct calculation of the each edge label. 

\subsection{Direct calculation of edge labels}
We begin by making a number of important additional definitions.  For a given node $q(x)$ and a given orbital index $i \in [1,n_o]$,  we define
 \begin{align}
 \label{OO}
 O_{<i}^{\text{incl}}(x)&:=\left\{j; \:\:1\le j<i, \:\chi_j \in \text{orb}(x)\right\},\\
 O_{<i}^{\text{excl}}(x)&:=\left\{j; \:\:1\le j<i, \:\chi_j \notin \text{orb}(x)\right\}, 
 \end{align}
 to denote the set containing the orbitals below the orbital $i$ that are included in the configuration of $\ket{x}$ (see (\ref{def:spec})), and the set of orbitals below orbital $i$ that are excluded from the configuration of $\ket{x}$, respectively. Note that orbital $i$ itself need not be in the configuration of $\ket{x}$. 
 Also let 
 \begin{align}
 n^{\text{incl}}_{i}(x)&:=|O_{<i}^{\text{incl}}(x)|,\\
 n^{\text{excl}}_i(x)&:=|O_{<i}^{\text{excl}}(x)|,  
 \end{align}
 denote the number of orbitals below orbital $i$ that are included and excluded in the configuration of $\ket{x}$. 
The following equivalent relations hold for $n^{\text{incl}}$ and $n^{\text{excl}}$: 
\begin{align}
n^{\text{incl}}_i(x)=\text{Ham}\left(x[i]\right), \\
n^{\text{excl}}_i(x)=i-1-\text{Ham}\left(x[i]\right), 
\end{align}
where 
 \begin{align}
 x[i]:=x\mod 2^{i}. 
 \end{align}
  and Ham is the Hamming function. 
  
Finally let 
\begin{align}
O_{>i}^{\text{incl}}(x):=\left\{j; \:\:i< j\leq n_o, \:\chi_j \in \text{orb}(x)\right\}
\end{align} 
be the set of orbitals \emph{above} orbital $i$ that are included in the configuration of $\ket{x}$.  We first consider the case of edges connecting two nodes that differ in one orbital excitation, {\it i.e.} states whose configurations differ in only one orbital. Afterwards, we consider the edges connecting two nodes that differ in two orbital excitations.  
\subsection{Single Excitations}

Consider  two states $\ket{x}$ and $\ket{y}$ that differ by the exchange of one occupied orbital, where $y>x$.  Let $u(x,y)$  denote the orbital in $\text{orb}(y)$ that is excluded from $\text{orb}(x)$, and $d(x,y)$  denote the orbital in $\text{orb}(x)$ that is excluded from $\text{orb}(y)$. As we saw earlier, $u(x,y)>d(x,y)$, where we drop the arguments of $u \equiv u(x,y)$ and $d \equiv d(x,y)$  wherever the context is clear (We have chosen $u$ standing for `up' and $d$ standing for `down'.).   
We denote the edge labels connecting such nodes as $e^{(1)}_{x\rightarrow y}$ and $e^{(1)}_{y\rightarrow x}$. We have
\begin{align}
e^{(1)}_{x\rightarrow y}&=\sum_{i\in O_{>u}^{\text{incl}}(x)}n_{i}^{\text{excl}} (x)+n_{u}^{\text{incl}} (x)\:n_{u}^{\text{excl}} (x)+1.\\
\end{align}
To see why, note that the first term in the right hand side sums the number of nodes below $q(y)$ that lack an orbital of $\text{orb}(x)$ higher than $u$, and instead include a lower orbital that $\text{orb}(x)$ lacks. The sec on term counts the nodes smaller than $y$ that lack and orbital of of $\text{orb}(x)$ lower than $u$ while, instead, having a lower orbital that $\text{orb}(x)$ lacks. The last term simply counts the node $q(x)$ itself.  

Similarly, 
\begin{align}
e^{(1)}_{y\rightarrow x}&= \sum_{i\in O_{>u}^{\text{incl}}(x)}n_i^{\text{excl}} (x)+n_d^{\text{excl}} (x)+1. 
\end{align}
Again, the first term counts the number of nodes below $q(x)$ that lack an orbital of $\text{orb}(y)$ higher than $u$, and instead include a lower orbital that $\text{orb}(y)$ lacks, as well, simply because $\ket{x}$ and $\ket{y}$ share the exact same orbitals higher than $u$. The second term counts the remaining number of states lower than $q(x)$ that differ with $\ket{y}$ in one orbital. They are those states that lack orbital $u$ and instead have an orbital less than $d$, as otherwise they would have had an orbital higher than $d$ that $\ket{x}$ lacks and thus their node would have been higher than $q(x)$. Again, the last term simply counts the node $q(x)$ itself.

\subsection{Pair Excitations}
Consider two states $\ket{x}$ and $\ket{y}$, where $x<y$,  that differ by two occupied orbitals in their configurations. Let $u_h(x,y)$,  $u_\ell(x,y)$ denote the orbitals in $\text{orb}(y)$ that are excluded in $\text{orb}(x)$,  where $u_h(x,y)>u_\ell(x,y)$. Also let $d_h(x,y)$ and  $d_\ell(x,y)$  denote the orbitals in $\text{orb}(x)$ that are excluded in $\text{orb}(y)$,  where $d_h(x,y)>d_\ell(x,y)$.  As we assumed $x<y$, we also have 
\begin{align}
u_h(x,y)>d_h(x,y), \:\: u_\ell(x,y)>d_\ell(x,y).
\end{align}
Again, we drop the arguments of $u_h \equiv u_h(x,y)$ etc., in our adopted notation wherever the context is clear. 

We denote the edge labels connecting such nodes as $e^{(2)}_{x\rightarrow y}$ and $e^{(2)}_{y\rightarrow x}$. In this case, we have 
\begin{align}
e^{(2)}_{x\rightarrow y}&=\sum_{i\in O_{>u_h}^{\text{incl}}(x)}n_{i}^{\text{incl}} (x) {n_{i}^{\text{excl}} (x) \choose 2}  +{n_{u_h}^{\text{incl}} (x)\choose 2} {n_{u_h}^{\text{excl}} (x)  \choose 2}+1.  
\end{align}
The reason why is as follows: The first term counts all the nodes lower than $q(y)$ that lack at least one orbital higher than $u$ that $\text{orb}(x)$ contains, but have another lower orbital instead of it that is not in $\text{orb}(x)$. Each term in that sum of occupied orbitals $i$ in $O_{>u_h}^{\text{incl}}(x)$  (so that $i>u_h$) is equal the number of ways to choose another occupied orbital of $x$, beside the orbital $i$,   times the number of ways two lower unoccupied orbitals excluded in $\text{orb}(x)$ can be chosen. The occupied orbitals are the ones excluded  and the chosen unoccupied orbitals  are instead the ones  included in the configurations of the node that is being thus counted.  
The second term similarly counts all nodes that differ in two occupied orbitals lower than $u_h$. The last term, as always, counts the node $q(x)$ itself. 

Similarly, for the edge in the other direction, we have 
\begin{align}
e^{(2)}_{y\rightarrow x}&= \sum_{i\in O_{>u_h}^{\text{incl}}(x)}n_{i}^{\text{incl}} (x) {n_{i}^{\text{excl}} (x) \choose 2}  +{n_{d_h}^{\text{excl}} (x) \choose 2} + \:n_{d_\ell}^{\text{excl}} (x)+1.
\end{align}
To see how, again note that the first term on the right hand side is the same as in the previous case, since $\ket{x}$ and $\ket{y}$ have the exact same orbitals higher than $u_h$. The second term counts the number of nodes that differ in the two occupied orbitals of $\ket{y}$, namely $u_h$ and $u_d$, having instead two other orbitals lower than $d_h$.  States having new orbital higher than $d_h$ have nodes larger than $q(x)$.  The third term on the right hand side counts the states that also lack both  orbitals $u_h$ and $u_d$, but instead have the orbital $d_h$ included. Their other included orbital has to be lower than $d_\ell$ now since otherwise the node of that state will be larger than $q(x)$. The last term once again counts the state $\ket{x}$ itself. 

\subsection{Summary of Edge Labels}
To summarize, we have the following formulas for the edge labels:
\begin{align}
e^{(1)}_{x\rightarrow y}&=R_1(x,y;u)+n_u^{\text{incl}} (x)\:n_u^{\text{excl}} (x)+1. \label{eq:e1xy} \\
e^{(1)}_{y\rightarrow x}&=R_1(x,y;r_1)+n_d^{\text{excl}} (x)+1, \label{eq:e1yx} \\
e^{(2)}_{x\rightarrow y}&= R_2(x,y;u_h)+{n^{\text{incl}}_{u_h}(x)\choose 2} {n^{\text{excl}}_{u_h}(x) \choose 2}+1, \label{eq:e2xy} \\
e^{(2)}_{y\rightarrow x}&=R_2(x,y;u_h)+{n^{\text{excl}}_{d_h}(x) \choose 2} +\:n^{\text{excl}}_{d_\ell}(x)  +1, \label{eq:e2yx} 
\end{align}
where 
\begin{align}
R_1(x,y;u)&:=\sum_{i\in S_{>u}^{\text{incl}}(x)}n_i^{\text{excl}} (x), \label{def:R1}\\
R_2(x,y;u_h)&=\sum_{i\in O_{>u_h}^{\text{incl}}(x)}n_{i}^{\text{incl}} (x) {n_{i}^{\text{excl}} (x) \choose 2}. \label{def:R2}  
\end{align}
Next, we show that for any trio of states $\ket{x}, \ket{y}, \ket{z}$, where $x<y<z$ whose corresponding nodes on the graph are connected, at least two of the edge labels must always be unequal. The special structure of the graph based on the formulas derived here is ultimately due to the spacial features  feature of the CI-matrix in the Coupled Configuration approximation. It ensures that  decomposing the overall Hamiltonian into matrices who have zero elements everywhere except for basis states that are connected with a particular edge label pair $\hat{e}_{x,y}=(e_{x\rightarrow y}, e_{y\rightarrow x})$ will always be one-sparse and thus directly siimulable in one go at every step of the Trotterization process.   

\subsection{Main Theorem}
The quantity $R_1(x,y;u)$  depends directly on the set $O_{>u(x,y)}^{\text{incl}}(x)$. Our strategy is to consider three separate cases based on how the sets $O_{>u(x,y)}^{\text{incl}}(x)$ and $O_{>u'(y,z)}^{\text{incl}}(y)$ related to each other. We initially focus on states that differ in only a single excitation. The case of pair excitations follows a similar line of argumentation.   

\subsubsection{Single Excitations}

{\bf Case a:}
If $O_{>u(x,y)}^{\text{incl}}(x) \subset O_{>u'(y,z)}^{\text{incl}}(y)$, then the following inequalities hold: 
\begin{align}
n_{d'(y,z)}^{\text{excl}} (y)&> n_{d(x,y)}^{\text{excl}} (x)-n_{u(x,y)}^{\text{excl}}(x) \label{eq:n01}\\
R_1(y,z;u') &\geq  R_1(x,y;u)+n_{u(x,y)}^{\text{excl}} (x). \label{eq:ry1}
\end{align}
By definition $u(x,y)>d(x,y)$. The inequality~(\ref{eq:n01}) follows as $n_{d'(y,z)}^{\text{excl}} (y)$ is a positive integer.      
Inequality~(\ref{eq:ry1}) holds as  $u(x,y) \in O^{\text{incl}}_{>u'}(y)$ in this case, and,  therefore,  the sum in~(\ref{def:R1}) for $R_1(y,z;u')$ contains (at least) an extra term $n_{u(x,y)}^{\text{excl}} (x)$ compared  to $R_1(x,y;u)$. 
Adding the two inequalities~(\ref{eq:n01}) and (\ref{eq:ry1}) leads to 
\begin{align}
R_1(y,z;u'_1) + n^{\text{excl}}_{d'(y,z)} (y) >  R_1(x,y;u_1)+n^{\text{excl}}_{d(x,y)} (x),   
\end{align}
or equivalently, by Eq.~(\ref{eq:e1yx}),  to 
$$
e^{(1)}_{z\rightarrow y}>e^{(1)}_{y\rightarrow x}.
$$
\newline

{\bf Case b:} If $O_{>u'(y,z)}^{\text{incl}}(y)  \subset  O_{>u(x,y)}^{\text{incl}}(x)$, then there must be an orbital index $s>u(x,y)$ such that $s\in O^{\text{incl}}_{>u}(x)$ but $s \notin O^{\text{incl}}_{>u'}(x)$, adding $n^{\text{excl}}_{s}(x)$ to the sum in~(\ref{def:R1}) of $R_1(x,y;u)$ but not to $R_1(y,z;u')$:
\begin{align}
\label{eq:rleq}
R_1(y,z;u') &\leq R_1(x,y;u)-n^{\text{excl}}_{s}(x). 
\end{align}
From this we can derive a second inequality: 
\begin{align}
\label{eq:rleq0}
R_1(y,z;u') &\leq R_1(x,y;u)-n^{\text{excl}}_{d'(y,z)}(y).
\end{align}
To see this, note that if $d(y,z)<s$ then $n^{\text{excl}}_{d'(y,z)}(y)<n^{\text{excl}}_{s}(x)$ from which the inequality follows. If, however,  $d'(y,z)\geq s$ then the argument above about $s$ also holds for $d'(y,z)$, namely that $d'(y,z) \in O^{\text{incl}}_{>u}(x)$  but  $d'(y,z) \notin O^{\text{incl}}_{>u'}(y)$, again leading to~(\ref{eq:rleq0}).
We immediately conclude that 
$$
R_1(y,z;u')+n^{\text{excl}}_{d'(y,z)}(y) \leq R_1(x,y;u) \leq R_1(x,y;u)+n^{\text{excl}}_{d(x,y)}(x)
$$
or equivalently, from Eq.~(\ref{eq:e1xy}), that 
$$
e^{(1)}_{z\rightarrow y}\leq e^{(1)}_{y\rightarrow x}.
$$
The equality is reached only when 
\begin{align}
\begin{cases}
n^{\text{excl}}_{d(x,y)}(x)=0,\label{n0dx=0}\\
O_{>u(x,y)}^{\text{incl}}(x)\backslash O_{>u'(y,z)}^{\text{incl}}(y)=\left\{d'(y,z)\right\}. 
\end{cases}
\end{align}
Under these conditions we have
\begin{align}
R_1(x,y;u)&=R_1(y,z;u')+n^{\text{excl}}_{u'(y,z)}(y), \label{eq:R1R1n}\\
n^{\text{incl}}_{u'(y,z)}(y)&= n^{\text{incl}}_{u(x,y)}(x)+1. \label{eq:n1+1}
\end{align}
The relation~(\ref{eq:R1R1n}) follows directly from the definition of $R_1$ in~(\ref{def:R1}) under the condition~(\ref{n0dx=0}). Eq.~(\ref{eq:n1+1}) holds because the configurations of $\ket{y}$ includes  at least one extra occupied orbital  compared to the configuration of $\ket{x}$, namely the orbital $\chi_{d'(y,z)}$ that is counted in $O_{<u'(y,z)}^{\text{incl}}(x)$ among the occupied orbitals in the state $\ket{y}$  while it is not counted in $O_{<u(x,y)}^{\text{incl}}(x)$ because of~(\ref{n0dx=0}).  As 
$$n^{\text{excl}}_{u'(y,z)}(y)> n^{\text{excl}}_{u(x,y)}(x)$$ 

 we conclude that 
\begin{align}
\label{eq:R1R1n+1}
R_1(y,z;u')+n^{\text{excl}}_{u'(y,z)}(y)\:n^{\text{incl}}_{u'(y,z)}(y)&= R_1(y,z;u')+n^{\text{excl}}_{u'(y,z)}(y)\:n^{\text{incl}}_{u(x,y)}(x)+n^{\text{excl}}_{u'(y,z)}(y)\\
&> R_1(x,y;u)+n^{\text{excl}}_{u(x,y)}(x)\:n^{\text{incl}}_{u(x,y)}(x).
\end{align}
which in turn implies 
$$
e^{(1)}_{y\rightarrow z}> e^{(1)}_{x\rightarrow y}.
$$
\newline
{\bf Case c:} 
If $O_{>u'(y,z)}^{\text{incl}}(y)  = O_{>u(x,y)}^{\text{incl}}(x)$, then the following hold: 
\begin{align}
R_1(y,z;u')=R_1(x,y;u),\\
n^{\text{incl}}_{u'(y,z)}(y)= n^{\text{incl}}_{u(x,y)}(x),\label{eq:n1c}\\
n^{\text{excl}}_{u'(y,z)}(y)> n^{\text{excl}}_{u(x,y)}(x).\label{eq:n0c}
\end{align}
Inequality~(\ref{eq:n0c}) holds because the orbital $\chi_{d(x,y)}$ is excluded from the string $y$.  Equation~(\ref{eq:n1c}) holds because the orbital $\chi_{u(x,y)}$ is included instead in the string $y$. From this set of inequalities we get 
\begin{align}
R_1(y,z;u')+n^{\text{incl}}_{u'(y,z)}(y) n^{\text{excl}}_{u'(y,z)}(y)> R_1(x,y;u)+n^{\text{incl}}_{u(x,y)}(x) n^{\text{excl}}_{u(x,y)}(x).
\end{align}
Comparing with Eq.~(\ref{eq:e1xy}), we conclude that 
$$
e^{(1)}_{y\rightarrow z}>e^{(1)}_{x\rightarrow y}.
$$
We thus have shown that in none of the possible cases can both the labels for the two pairs be equal. 

\subsubsection{Pair Excitations}

{\bf Case a:}
If $O_{>u_h(x,y)}^{\text{incl}}(x) \subset O_{>u'_h(y,z)}^{\text{incl}}(y)$, then we have: 
\begin{align}
n_{d'_h(y,z)}^{\text{excl}} (y)&> n_{d_h(x,y)}^{\text{excl}} (x)-n_{u_h(x,y)}^{\text{excl}}(x) \label{eq:n02}\\
R_2(y,z;u'_h) &\geq  R_2(x,y;u_h)+{n_{u_h(x,y)}^{\text{excl}} (x) \choose 2}. \label{eq:ry2}
\end{align}
Again, by definition, we have $u_h(x,y)>d_h(x,y)$. The inequality~(\ref{eq:n02}) holds because $n_{d'_h(y,z)}^{\text{excl}} (y)$ is a positive integer.      
Inequality~(\ref{eq:ry2}) holds as  $u_h(x,y) \in O^{\text{incl}}_{>u'_h}(y)$ in this case, and thus the sum in~(\ref{def:R2}) for $R_2(y,z;u'_h)$ contains (at least) an extra term ${n_{u_h(x,y)}^{\text{excl}} (x) \choose 2}$ compared  to $R_2(x,y;u_h)$. 
By adding the two inequalities~(\ref{eq:n02}) and (\ref{eq:ry2}) we get  
\begin{align}
R_2(y,z;u'_h) + {n^{\text{excl}}_{d'_h(y,z)} (y) \choose 2}>  R_2(x,y;u_h)+{n^{\text{excl}}_{d_h(x,y)} (x) \choose 2},   
\end{align}
which, by Eq.~(\ref{eq:e2yx}), amounts to 
$$
e^{(2)}_{z\rightarrow y}>e^{(2)}_{y\rightarrow x}.
$$
\newline

{\bf Case b:} If $O_{>u'_h(y,z)}^{\text{incl}}(y)  \subset  O_{>u_h(x,y)}^{\text{incl}}(x)$, then there must be an orbital index $s>u_h(x,y)$ such that $s\in O^{\text{incl}}_{>u_h}(x)$ but $s \notin O^{\text{incl}}_{>u'_h}(x)$, adding $n^{\text{excl}}_{s}(x)$ to the sum in~(\ref{def:R2}) of $R_2(x,y;u_h)$ but not to $R_2(y,z;u'_h)$:
\begin{align}
\label{eq:rleq2}
R_2(y,z;u'_h) &\leq R_2(x,y;u_h)-{n^{\text{excl}}_{s}(x) \choose 2}. 
\end{align}
From this we can derive a second inequality: 
\begin{align}
\label{eq:rleq02}
R_2(y,z;u'_h) &\leq R_2(x,y;u_h)-{n^{\text{excl}}_{d'_h(y,z)}(y) \choose 2},   
\end{align}
and 
\begin{align}
\label{eq:rleq22}
R_2(y,z;u'_h) + {n^{\text{excl}}_{d'_h(y,z)}(y) \choose 2} &\leq R_2(x,y;u_h)\leq  R_2(x,y;u_h)+ {n^{\text{excl}}_{d_h(x,y)}(x) \choose 2}.
\end{align}
We consider the following two cases separately. 

{\bf case b1}. If $n^{\text{excl}}_{d'_h(y,z)}(y)< n^{\text{excl}}_{d_h(x,y)}(x)$, then we have 
\begin{align}
{n^{\text{excl}}_{d_h(x,y)}(x) \choose 2}-{n^{\text{excl}}_{d'_h(y,z)}(y) \choose 2} > n^{\text{excl}}_{d'_\ell(y,z)}(y)-n^{\text{excl}}_{d_\ell(x,y)}(x), 
\end{align}
since $n^{\text{excl}}_{d_\ell(x,y)}(x)\le n^{\text{excl}}_{d_h(x,y)}(x)$ and $n^{\text{excl}}_{d'_\ell(y,z)}(y)\le n^{\text{excl}}_{d'_h(y,z)}(y)$.  From~(\ref{eq:rleq22}) and~(\ref{eq:e2yx}) it directly follows that 
$$
e^{(2)}_{z\rightarrow y}<e^{(2)}_{y\rightarrow x}.
$$

{\bf case b2}. If $n^{\text{excl}}_{d'_h(y,z)}(y)\ge n^{\text{excl}}_{d_h(x,y)}(x)$,  we note that 
\begin{align}
\label{eq:sss2}
R_2(x,y;u_h) -R_2(y,z;u'_h)\ge {n^{\text{excl}}_{s}(x) \choose 2},  
\end{align}
 and 
\begin{align}
\label{eq:sss3}
{n^{\text{excl}}_{s}(x) \choose 2} \ge {n^{\text{excl}}_{d'_h(y,z)}(y) \choose 2}+n^{\text{excl}}_{d'_\ell(y,z)}(y) -{n^{\text{excl}}_{d_h(x,y)}(x) \choose 2}-n^{\text{excl}}_{d_\ell(x,y)}(x), 
\end{align}
or equivalently, 
$$
e^{(2)}_{z\rightarrow y}\leq e^{(2)}_{y\rightarrow x}.
$$
Equality holds simultaneously in~(\ref{eq:rleq22}),~(\ref{eq:sss2}) and~(\ref{eq:sss3}) only if 
\begin{align}
\begin{cases}
n^{\text{excl}}_{d_h(x,y)}(x)=n^{\text{excl}}_{d_\ell(x,y)}(x)=0\\
O_{>u_h(x,y)}^{\text{incl}}(x)\backslash O_{>u'_h(y,z)}^{\text{incl}}(y)=\left\{d'_h(y,z)\right\}.
\end{cases}
\end{align} 
By arguments similar to those applied to the single-excitation case in Eqs.~(\ref{eq:R1R1n})-(\ref{eq:R1R1n+1}), we conclude that 
\begin{align}
R_2(x,y;u_h)&=R_2(y,z;u'_h)+{n^{\text{excl}}_{u'_h(y,z)}(y) \choose 2}, \label{eq:R2R2n}\\
n^{\text{incl}}_{u'_h(y,z)}(y)&= n^{\text{incl}}_{u_h(x,y)}(x)+1. \label{eq:n1+2}
\end{align}
and consequently 
\begin{align}
R_2(y,z;u'_h)+{n^{\text{excl}}_{u'_h(y,z)}(y) \choose 2} \:{n^{\text{incl}}_{u'_h(y,z)}(y) 
\choose 2}> R_2(x,y;u_h)+{n^{\text{excl}}_{u_h(x,y)}(x) \choose 2}\:{n^{\text{incl}}_{u_h(x,y)}(x) \choose 2}.
\end{align}
which in turn implies 
$$
e^{(2)}_{y\rightarrow z}> e^{(2)}_{x\rightarrow y}.
$$
\newline

{\bf Case c:} 
If $O_{>u'_h(y,z)}^{\text{incl}}(y)  = O_{>u_h(x,y)}^{\text{incl}}(x)$, the argument is similar to the one we used for the case of single excitations. We conclude for similar reasons that the following hold: 
\begin{align}
R_2(y,z;u'_h)=R_2(x,y;u_h),\\
n^{\text{incl}}_{u'_h(y,z)}(y)= n^{\text{incl}}_{u_h(x,y)}(x),\label{eq:n1c2}\\
n^{\text{excl}}_{u'_h(y,z)}(y)> n^{\text{excl}}_{u_h(x,y)}(x).\label{eq:n0c2}
\end{align}
This set of equations lead to 
\begin{align}
R_2(y,z;u'_h)+{n^{\text{incl}}_{u'_h(y,z)}(y) \choose 2} {n^{\text{excl}}_{u'_h(y,z)}(y) \choose 2}> R_2(x,y;u_h)+{n^{\text{incl}}_{u_h(x,y)}(x) \choose 2} {n^{\text{excl}}_{u_h(x,y)}(x) \choose 2}.
\end{align}
Comparing with Eq.~(\ref{eq:e2xy}), we conclude that 
$$
e^{(2)}_{y\rightarrow z}>e^{(2)}_{x\rightarrow y}.
$$
We have shown that none of the labels for the two pairs can be equal in the case of double excitations either, and this concludes the proof.

\end{document}